\documentclass[submission,copyright,creativecommons]{eptcs}
\providecommand{\event}{FROM 2023} 

\usepackage{iftex}
\usepackage{alltt}
\ifpdf
  \usepackage{underscore}         
  \usepackage[T1]{fontenc}        
\else
  \usepackage{breakurl}           
\fi

\usepackage{pifont}
\usepackage{algpseudocode}
\newcommand{\cmark}{\ding{51}}%
\newcommand{\xmark}{\ding{55}}%

\title{Identifying Vulnerabilities in Smart Contracts \\ using Interval Analysis}
\author{{\c S}tefan-Claudiu Susan
\institute{Alexandru Ioan Cuza University of Iași,\\ Department of Computer Science\\
Iași, România}
\email{claudiu\_susan@yahoo.com}
\and
Andrei Arusoaie
\institute{Alexandru Ioan Cuza University of Iași,\\ Department of Computer Science\\
Iași, România}
\email{andrei.arusoaie@uaic.ro} 
}
\def\titlerunning{Identifying vulnerabilities in Smart Contracts using Interval Analysis}
\def\authorrunning{Ștefan-Claudiu Susan \& Andrei Arusoaie}
\begin{document}
\maketitle

\begin{abstract}


This paper serves as a progress report on our research, specifically focusing on utilizing interval analysis, an existing static analysis method, for detecting vulnerabilities in smart contracts. We present a selection of motivating examples featuring vulnerable smart contracts and share the results from our experiments conducted with various existing detection tools. Our findings reveal that these tools were unable to detect the vulnerabilities in our examples. To enhance detection capabilities, we implement interval analysis on top of Slither~\cite{slither}, an existing detection tool, and demonstrate its effectiveness in identifying certain vulnerabilities that other tools fail to detect.
\end{abstract}

\section{Introduction}

The term ``smart contract'' was originally used to describe automated legal contracts, whose content cannot be negotiated or changed. Nowadays, the term is most commonly known as programs that are executed by special nodes in a decentralised network or a blockchain. Indeed, the blockchain technology captures the initial meaning of the term: contracts are encoded as an immutable piece of code, and the terms of the contract are predetermined and automatically enforced by the contract itself.

This immutability property also implies more effort on the contract developers side: they have to be very careful about what gets deployed because that code is (1) public and anyone can see it and (2) it cannot be changed/updated as an ordinary program. Ethereum\footnote{Vitalik Buterin, \emph{A Next-Generation Smart Contract and Decentralized Application Platform}, 2014, URL: \url{https://ethereum.org/en/whitepaper/}} is a very popular blockchain platform which has been affected by the most significant attacks based on vulnerabilities in the deployed code. For instance, ``The DAO attack''~\footnote{David Siegel, 2016: Understanding The DAO Attack. URL:\url{https://www.coindesk.com/learn/2016/06/25/understanding-
the-dao-attack/}} was based on the fact that a smart contract could be interrupted in the middle of its execution and then called again. This is known as the \emph{reentrancy} vulnerability. An attacker noticed that in a withdrawal function of the smart contract, the transfer of digital assets was performed before updating the balance (i.e., decrementing the balance with the withdrawn amount) of a contract party.
The attacker first deposited cryptocurrency into the smart contract. Then, by creating a scenario where the withdrawal function called itself just before updating its own balance, the attacker managed to drain the funds of the smart contract as long as its balance exceeded the amount withdrawn.


Such mistakes are unfortunate and researchers and practitioners started to propose methods and tools for detecting them. 
For example, Slither~\cite{slither} is an easy to use static analysis tool for smart contracts written in Solidity~\footnote{Solidity, version 0.8.20: \url{https://docs.soliditylang.org/en/v0.8.20/control-structures.html}}; Mythril\footnote{Mythril docs: \url{https://mythril-classic.readthedocs.io/en/develop/}} is a security analysis tool for EVM bytecode based on symbolic execution; Solhint\footnote{Solhint official website: \url{https://protofire.github.io/solhint/}} is a linter for Solidity code. The list of tools is long and it was explored in various papers (e.g.,~\cite{10.1145/3282373.3282419,10.1145/3464421,10.3389/fbloc.2022.814977,10.1007/978-3-662-54455-68,10.1007/978-3-319-89722-610}).
These tools are indeed very useful, but in the same time they are not perfect and they can fail to detect problematic situations in smart contracts code.

In this paper we provide several examples of smart contracts which contain vulnerabilities and we find that vulnerability detection tools are not as precise as we expect, and they fail to detect vulnerabilities in our examples. We attempt to enhance one of them (Slither) with an existing static analysis method called interval analysis. This method allows us to better approximate the values interval for each program variable. Based on the experiments that we performed, interval analysis proves to be very useful in detecting problematic situations in smart contracts. For example, integer division in Solidity ignores the reminder. In a situation where an amount of cryptocurrency must be divided and transferred to a number of recipients, a division where the remainder is ignored could lead to funds that remain locked in the smart contract. Another example is related to uninitialised variables: such variables are initialised with default values and it may be the case that the default value is not suitable for the purpose of that variable. 
By keeping track of all the possible values for each program variable, interval analysis allows us to signal such situations in smart contracts.\\[-5ex]
\paragraph{Summary of contributions.}
\begin{enumerate}
\setlength\itemsep{0em}
    \item We provide several examples of vulnerable smart contracts, in which the vulnerabilities prove to be challenging to detect using state-of-the-art detection tools.
    \item We implement an existing analysis technique called interval analysis on top of Slither.
    \item We evaluate our implementation.\\[-5ex]
\end{enumerate}
\paragraph{Paper organisation.}
In Section~\ref{sec:vulnerabilities} we present several examples of smart contract vulnerabilities. In Section~\ref{sec:existingtools} we show how state-of-the-art tools behave on these examples. The interval analysis technique is presented in Section~\ref{sec:intervalanalysis} together with our implementation. We  conclude in Section~\ref{sec:conclusions}.

\section{Vulnerabilities in Smart Contracts}
\label{sec:vulnerabilities}
This section contains several examples of smart contracts vulnerabilities written in Solidity. These were selected from a larger taxonomy~\cite{10.3389/fbloc.2022.814977}. The whole classification includes 55 vulnerabilities split among 10 categories. Both literature and existing community taxonomies were taken into account when selecting these defects. We selected these vulnerabilities because state of the art tools are not able to detect most of them and could be detectable using interval analysis. 

\subsection{Tautologies or Contradictions in \texttt{assert} or \texttt{require} Statements}
The Solidity statements \texttt{assert} and \texttt{require} are typically used to validate boolean conditions. According to the Solidity documentation\footnote{Solidity docs: \url{https://docs.soliditylang.org/en/v0.8.20/control-structures.html}}, \texttt{assert} is meant for checking internal errors, while \texttt{require} should be used to test conditions that cannot be determined until runtime. Both statements throw exceptions and revert the corresponding transactions. In their intended use, the conditions in \texttt{assert} should never be false as it signals contract level errors while the conditions in \texttt{require} can be false as they signal input errors. No matter what level of error a statement specifies, it is an issue if the conditions that they contain are tautologies or contradictions. These make the statement useless in the case of tautologies and make the transaction impossible to complete in the case of contradictions as illustrated by the following code: 

{\small
\begin{alltt}
1:  function notGonnaExecute(uint parameter) external pure returns(uint)
2:  \{
3:      require(parameter<0); // uint cannot be < 0
4:      return parameter;
5:  \}
\end{alltt}
}

{\small
\begin{alltt}
1:  function uselessAssertUint(uint parameter) external pure returns(uint)
2:  \{
3:      require(parameter>=0); // uint is always >= 0
4:      return parameter;
5:  \}
\end{alltt}
}
\subsection{Division by Zero}
This is a classic arithmetic issue that is common among most programming languages. The Solidity compiler does not allow direct division by zero.
However, the compiler cannot detect situations when the denominator could evaluate to zero.
The following code snippet contains an example. The length of the recipients array is not checked before computing the amount that should be sent to each recipient~(line~4):
{\small
\begin{alltt}
1:  function split(address[] calldata recipients) external payable
2:  \{
3:      require(msg.value > 0,"Please provide currency to be split among recipients");
4:      uint amount = msg.value / recipients.length; // problem here if length is 0
5:      for(uint index = 0; index < recipients.length; index++)
6:      \{
7:          (bool success,) = payable(recipients[index]).call{value:amount}("");
8:          require(success,"Could not send ether to recipient");        
9:      \}
10: \}
\end{alltt}
}
\subsection{Integer Division Remainder}
This is another arithmetic issue that is common among many programming languages. Solidity performs integer division which means that the result of the division operation is truncated. This could lead to situations where ignoring the remainder of the division could lead to logic errors.
The snippet below contains an example: if the provided amount does not exactly divide by the number of recipients then that amount of cryptocurrency could remain locked in the contract.

{\small
\begin{alltt}
1:  function split(address[] calldata recipients) external payable
2:  \{
3:       require(recipients.length > 0,"Empty recipients list");
4:       uint amountPerRecipient = msg.value / recipients.length; // remainder ???
5:       require(amountPerRecipient > 0,"Amount must be positive");
6:       for(uint index = 0; index < recipients.length; index++)
7:       \{
8:           payable(recipients[index]).transfer(amountPerRecipient);
9:       \}
10:  \}
\end{alltt}
}
\subsection{Uninitialised Variable}
Uninitialized variables could lead to logical errors or exceptions. If a variable is not initialised, there is a great chance that the default value assigned to the variable (according to its type) is not suitable for the purpose of that variable.
The following code contains an access modifier which relies on the \texttt{owner} state variable. 
The variable is \texttt{private}, and thus, it cannot be accessed or assigned outside the contract. 
Also, there is no explicit initialisation of \texttt{owner} within a constructor. 
This makes the variable stuck to the default value, and thus, all the functions marked with the \texttt{onlyOwner} modifier cannot be executed.

{\small
\begin{alltt}
1:    address private owner;
2:
3:    modifier onlyOwner() \{
4:        require(msg.sender == owner, "Only the owner of the contract has access");
5:        _;
6:    \}
\end{alltt}
}\vspace{-3ex}

\subsection{User Input Validation}
Parameter validation or ``sanitisation'' is a process that must be implemented at the beginning of every method. This ensures that the method will always execute as expected. End users should not be trusted to always provide valid parameters. If validation is missing and the end user is unaware, or worse, malicious, it could cause critical errors that produce unexpected results or halt contract execution all together.
The following example contains a getter method for an internal array. The user can provide an index that is not validated, thus having the possibility of going out of bounds.
{\small
\begin{alltt}
1:    uint256[] private _array= [10, 20, 30, 40, 50]; 
2:
3:    function getArrayElement(uint256 index) external view returns (uint256)
4:    \{
5:        return _array[index];
6:    \}
\end{alltt}
}\vspace{-3ex}

\subsection{Unmatched Type}
\label{sec:unmatchedtype}
In Solidity, enums are stored as unsigned integers. Thus, they can be compared and assigned with variables of type \texttt{uint}. Situations like these can become tricky since the value domain of an enum is likely to be much smaller than the value domain of unsigned integers. If a variable with a greater value than the range of the enum is assigned to an enum variable, than the transaction will be reverted. While it is true that reverting the transaction is considered safe, such situations signal a faulty logic in the contract code and it is preferable to be avoided.
{\small
\begin{alltt}
1:  contract UnmatchedType \{
2:      enum Options \{ Candidate1, Candidate2, Candidate3 \}
3:      mapping(address => Options) private _votes;
4:      mapping(Options => uint) private _votesCount;
5:      function vote(uint option) external \{
6:          _votes[msg.sender] = Options(option);
7:          _votesCount[Options(option)]++;
8:      \}
9:      function getStatisticsForOption(uint option) external view returns(uint) \{
10:         return _votesCount[Options(option)];
11:     \}
12: \}
\end{alltt}
}\vspace{-3ex}
\section{Detecting Vulnerabilities Using Dedicated Analysis Instruments}
\label{sec:existingtools}

This section briefly presents the results of some experiments that we performed. 
Basically, we used a few tools for analysing smart contracts in order to check how they behave on our examples presented in Section~\ref{sec:vulnerabilities}. The tools that we selected are presented below. It is worth noting that we picked tools which implement different techniques (e.g., static analysis, symbolic execution, linter). For a tool to be eligible for our study, it has to be open source, active and compatible with the latest version of Solidity.

Slither is a static analysis tool written in Python. It provides vulnerability detection and code optimization advice. It features many detectors that target different issues. Its analysis runtime is very low compared to the other tools. It analyses Solidity code by transforming the EVM bytecode into an intermediary representation called SlithIR. Being an open source project, it allows anyone to contribute and improve it, being the foundation for our implementation (discussed later in Section~\ref{sec:intervalanalysis}). Slither was able to detect uninitialized variables as well as trivial tautologies and contradictions in our examples.

Solhint is a linter for Solidity code. An open source project, it is able to detect possible vulnerabilities, optimization opportunities and abidance to style conventions. The tool also features a customisable set of detection rules that can be employed, along with predefined configurations. The user can define its own configurations and decide which issue wants to target. Unfortunately, Solhint was not able to detect any of the issues in our examples.

Remix\footnote{Remix Docs: \url{https://remix-ide.readthedocs.io/en/latest/}} also features a static analysis plugin. We were unable to find any information about the analysis process performed by this tool. Moreover, it was unable to detect any problems in our examples.

Mythril is a tool that leverages symbolic execution to simulate multiple runs of a contract's methods. It has a fairly long runtime compared to the others. We even encountered executions that took more than a few hours. Mythril was unable to detect any of the issues presented above.

In Table~\ref{tab:evaluationoftools}, we present a summary of the results that we obtained. The results indicate that nearly all tools fail to detect the vulnerabilities in our examples. This does not mean that these tools are not useful or very bad at signaling issues in smart contract code. The way we interpret these results is that these tools need to be enhanced with more powerful techniques that could increase their detection capabilities.\\[-4ex]
\begin{table}[t]
    \centering
    \begin{tabular}{|l|c|c|c|c|}
        \hline
         \textbf{Examples} & Slither & Solhint & Remix & Mythril \\
         \hline \hline
         Tautologies/Contradictions &  \cmark & \xmark & \xmark & \xmark\\
         Division by zero & \xmark & \xmark & \xmark& \xmark\\
         Integer division & \xmark & \xmark & \xmark& \xmark\\
         Uninitialised variable & \cmark & \xmark & \xmark& \xmark\\
         User input validation & \xmark & \xmark & \xmark& \xmark\\
         Unmatched type & \xmark & \xmark & \xmark& \xmark\\
         \hline
    \end{tabular}
    \caption{A summary of the evaluation of the tools when executed on our examples (Section~\ref{sec:vulnerabilities}).}
    \label{tab:evaluationoftools}
\end{table}

\section{Interval Analysis for Vulnerability Detection}
\label{sec:intervalanalysis}
\subsection{Interval Analysis}
Interval Analysis~\cite{IntervalAnalysis} is a static analysis technique that approximates the values interval for every variable in a program for a certain instruction. 
The technique is not limited to predicting the values interval of a variable, it can also be used to predict certain properties that can be derived from the value of the variable. For instance, instead of working with integer intervals, an analysis can target the parity of variables and work only with 2-valued intervals (even, odd).

We present interval analysis via the \texttt{Unmatched Type} example from Section~\ref{sec:unmatchedtype}. Moreover, we show how interval analysis can help us detect a problem in this example, more precisely in the \texttt{vote} function:

{\small
\begin{alltt}
5:      function vote(uint option) external \{
6:          _votes[msg.sender] = Options(option); //Statement 1
7:          _votesCount[Options(option)]++;       //Statement 2
8:      \}
\end{alltt}
}

\begin{table}[t]
    \centering
    \begin{tabular}{|l|c|c|c|}
        \hline
         \textit{Statements}  & \texttt{option} & \texttt{_votes[msg.sender]}\\
         \hline \hline
         1 &  [0, max] & [0, max] \\
         2 &  [0, max] & [0, 2]\\
         End & [0, max] & [0, 2]\\
         \hline
    \end{tabular}
    \caption{Interval analysis for the \texttt{vote} function.}
    \label{tab:intervalanalysisvote}
\end{table}

\noindent
The function registers the vote of an user and increases the total vote count for its option. The problem is at line 2: the input is of type \texttt{uint} and it could easily be outside the values range [0,1,2] of the enum.

Interval analysis provides an approximation of the values interval for every program variable at each program location.
In Table~\ref{tab:intervalanalysisvote} we show how these intervals are computed for our example. Each line of the table presents the intervals for the program variables (displayed in columns) \emph{before} the execution of each statement in the first column. 
For example, before the execution of \texttt{Statement 1}, we do not have any information about the \texttt{option} variable, so its range of values will correspond to the values domain for \texttt{uint}. 
For \texttt{_votes[msg.sender]}, the value interval changes before \texttt{Statement 2} in case of normal execution (otherwise, the transaction is reverted) to [0,2], that is, the only possible range for \texttt{Option}.
Interval analysis performs this calculation using the Worklist Algorithm, an algorithm which traverses the program control flow graph, and updates the intervals for these variables until a fixpoint is reached. This algorithm is shown in Section~\ref{sec:implementation}.

Recall that the problem we are trying to detect using interval analysis is a mismatch of domains between the variable assigned and the variable whose value is assigned. 
Since interval analysis computes the interval for \texttt{option} and \texttt{_votes[msg.sender]}, a close inspection of the difference between the intervals is sufficient to reveal the problem. A \texttt{require} statement that checks upfront the values for the \texttt{option} parameter would solve the problem. Also, our detection technique would not signal an issue.

\subsection{An Implementation of Interval Analysis on Top of Slither}
\label{sec:implementation}
We built our implementation using Python modules provided by Slither. These are the same modules that are used internally by Slither for its 
own detectors. During execution, Slither fills some of its internal data structures with useful information, such as contract CFG (control flow graph), an intermediary SSA (single statement assignment) representation of the code, and information about each variable (e.g., type, scope and name). We use the information in these data structures to implement interval analysis.

The Worklist Algorithm shown in Figure~\ref{fig:worklistalg} works by processing every edge in the contract CFG. 
These edges are added into a list (the "worklist"). 
It is an iterative algorithm that processes existing elements until the list is empty. 
When new information is added to the current state, new edges are also added to the worklist. 
The algorithm stops when no more new information can be discovered. 

\begin{figure}[ht]
\centering
\begin{minipage}{0.5\linewidth}
\begin{algorithmic}
\State $Values \gets InitialValues(G)$
\State $Worklist \gets RootSet(G)$
\While{$HasMoreNodes(Worklist)$}
\State $n_i \gets NextNode(Worklist)$
\While{$HasMoreEdges(n_i)$}
\State    $e \gets nextEdge(n_i)$
\State    $t \gets type(e)$
\State    $n_j \gets farNode(e,n_i)$
\State    $v'_j \gets F(t,v_i,v_j)$
    \If{$MonotonicChange(v'_j,v_j)$}
    \State $v_j \gets v'_j$
    \State $AddToWorklist(Worklist,n_j)$
\EndIf
\EndWhile
\EndWhile\\
return V
\end{algorithmic}
\end{minipage}
    \caption{The Worklist Algorithm.}
    \label{fig:worklistalg}
\end{figure}

We implemented a modular Worklist algorithm. Essential information such as extreme labels\footnote{The program nodes where the analysis begins.}, order function\footnote{A function that receives two elements of the same domain and determines the greater one.} and flow function\footnote{A function determining the edges in the flow graph or the reverse of those edges depending on the type of analysis.} are all provided as parameters to the Worklist algorithm. This allow us to perform multiple types of analysis using the same base implementation. 
Our implementation leverages the CFG provided by Slither 
to split the code of a function into multiple parts. Each node is then split even further into SlithIR SSA~\cite{cytron1986code}
lines that are analyzed individually. Along with basic types such as \texttt{uint} and \texttt{bool}, our implementation is able to model complex types such as arrays, mappings and structs.

We defined our own data type to encapsulate information about a variable such as type, scope, name and, most importantly, values interval. The program state is represented as a dictionary having variable names as keys and an object of our own defined type as values. Complex types are defined as recursive dictionaries, for example, the interval for a struct is modeled as a dictionary containing intervals for each of its fields or even other dictionaries if the structs are nested.\\[-5ex]

\paragraph{Current status.} Our implementation is now able to successfully analyze programs containing assignments and arithmetic expressions for both elementary and complex types. It takes into consideration state variables, function parameters, and local variables. 
We are now capable of detecting issues such as:
\begin{itemize}
\setlength\itemsep{-.2em}
    \item Arithmetic issues including \texttt{Division By Zero} and \texttt{Integer Division Remainder};
    \item Issues related to variables initialization;
    \item Issues related to parameter validation.
\end{itemize}~\vspace{-5ex}

\section{Conclusions}
\label{sec:conclusions}
In this paper we identified some vulnerabilities that are not handled by state of the art tools for smart contract analysis. These vulnerabilities vary in their severity, but no matter the impact, defects and potential errors should be identified as soon as possible. We attempt to improve Slither with a more powerful technique called interval analysis. We explain why this technique is a good fit for detecting these issues and how it could detect them. 
We built a custom interval analysis on top of Slither, leveraging the information that Slither already provides about a contract, its attributes, methods, method parameters, program flow and many more.  
Currently, our implementation detects vulnerabilities that other tools miss. 

\subsection{Future Work}
We are now handling only on a subset of expressions in the Solidity programming language, which covers expressions including integers, booleans, arrays, structures, and mappings. 
However, more elaborate work needs to be done to tackle addresses and operations over addresses, more complex loops or conditional statements, etc.
Right now, our code can be executed on every smart contract written in Solidity, but it will perform interval analysis only for the subset that we cover.

Intraprocedural analysis would significantly improve the precision of our analysis. 
An example of a vulnerability that we are not yet able to detect is \textit{Short Address}. This could be detected by monitoring the length attribute of the payload. Another example is \textit{Tautologies and Contradictions in Assert or Require Statements}: it could be detected by approximating the result of the boolean expression and checking if the interval contains only one value: \textit{true} for tautologies and \textit{false} for contradictions.

Being able to handle more complex conditional statements and loops would also be of great help in obtaining a more accurate monitoring of the program state by interpreting the semantics of boolean expressions. 
Once we identify multiple possible states based on conditional branches, we can leverage unifying techniques such as \textit{Trace Partitioning}.
Additionally, monitoring implicit state variables that are contract-level or function-level, like \texttt{balance} or \texttt{msg.sender}, would be beneficial in identifying balance-related issues and user interaction problems.\\[-5ex]

\bibliographystyle{eptcs}
\bibliography{generic}
\end{document}